\documentclass[showpacs,superscriptaddress,twocolumn,amsmath,aps]{revtex4}
\usepackage{graphicx}

\newcommand{\nl}{\nonumber \\}

\newcommand{\be}{\begin{equation}}
\newcommand{\ee}{\end{equation}}
\newcommand{\bea}{\begin{eqnarray}}
\newcommand{\eea}{\end{eqnarray}}

\newcommand{\Eq}[1]{Eq.\,(\ref{#1})}
\newcommand{\Eqs}[1]{Eqs.\,(\ref{#1})}

\newcommand{\la}{\langle}
\newcommand{\ra}{\rangle}
\newcommand{\dg}{\dagger}

\newcommand{\ep}{\epsilon}

\newcommand{\ti}{\tilde}

\begin{document}
\draft

\title{ Quantum Transport from the
        Perspective of Quantum Open Systems }
\author{Ping Cui}
\affiliation{Hefei National Laboratory for Physical Sciences at Microscale,
  University of Science and Technology of China, Hefei, China}
\affiliation{Department of Chemistry, Hong Kong University of Science and
         Technology, Kowloon, Hong Kong}

\author{Xin-Qi Li}
\email{xqli@red.semi.ac.cn}
\affiliation{State Key Laboratory for Superlattices and Microstructures,
         Institute of Semiconductors,
         Chinese Academy of Sciences, P.O.~Box 912, Beijing 100083, China}

\author{Jiushu Shao}
\affiliation{State Key Laboratory of Molecular Reaction Dynamics,
         Institute of Chemistry,
         Chinese Academy of Sciences, Beijing 100080, China }

\author{YiJing Yan}
\affiliation{Hefei National Laboratory for Physical Sciences at Microscale,
  University of Science and Technology of China, Hefei, China}
\affiliation{Department of Chemistry, Hong Kong University of Science and
         Technology, Kowloon, Hong Kong}

\date{\today}

\begin{abstract}
By viewing the non-equilibrium transport setup as a
quantum open system, we propose a reduced-density-matrix based
quantum transport formalism.
At the level of self-consistent Born approximation,
it can precisely account for the correlation between tunneling
and the system internal many-body interaction, leading to certain novel
behavior such as the non-equilibrium Kondo effect.
It also opens a new way to construct time-dependent density
functional theory for transport through large-scale complex
systems.
\end{abstract}
\maketitle

Conventionally, quantum transport and quantum dissipation are
treated with different approaches.
For instance, the former (in mesoscopic context) is usually described by the
Landauer-B\"uttiker theory and the non-equilibrium Green's function (nGF)
approach \cite{Dat95,Hau96},
whereas the latter by the reduced density matrix equation \cite{Gar00}.
Nevertheless, both are quantum open systems,
with either the non-equilibrium electron reservoirs (electrodes) or
the dissipative thermal bath as their environments, as schematically
shown in Fig.\ 1.
It is thus of great interest to develop a unified language to bridge them.

\begin{figure}
\begin{center}
\includegraphics*[width=8cm,height=4cm,keepaspectratio]{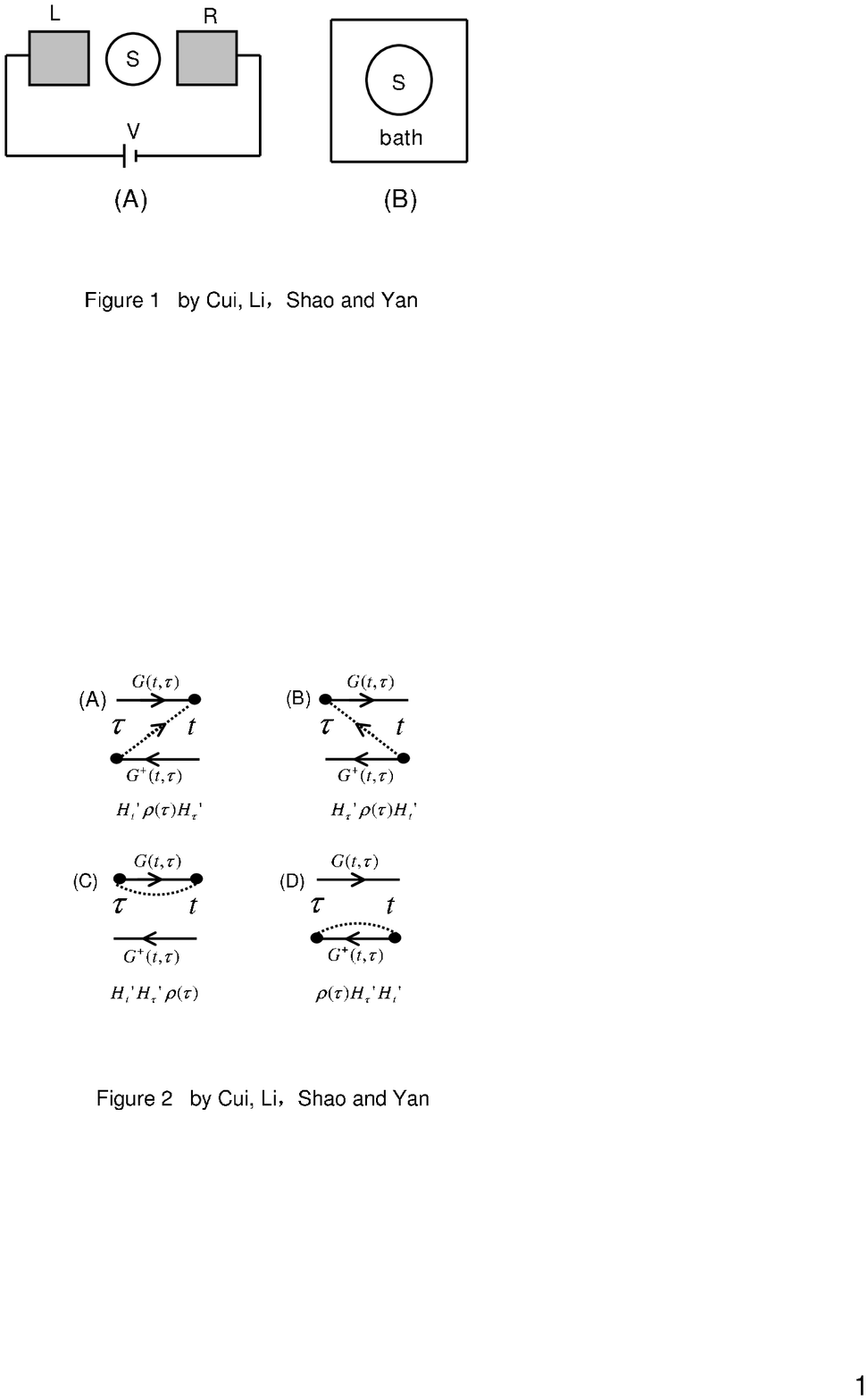}
\caption{A unified picture for (A) quantum transport and (B)
quantum dissipative open system, where the transport system is
regarded as the ``system of interest", and the electrodes as
``environmental baths". }
\end{center}
\end{figure}

This motivation can be historically dated back to the
phenomenological rate equation and quantum Bloch equation
approaches to transport \cite{Gla88,Gur96}.
There, either implicitly or explicitly, the electrodes
are viewed as dissipative reservoirs.
Along this line and based on our work
in quantum measurement of solid-state qubit \cite{Li04a,Li04b},
we developed recently a quantum master equation approach
to quantum transport \cite{Li04c}.
The reduced dynamics involved there was originally constructed
under the cumulant second-order expansion (Born approximation).
In this letter, we re-formalize it
in the spirit of self-consistent Born approximation (SCBA),
in order to make the formalism not only {\it convenient} but also
{\it accurate enough} in practice.
Moreover, by reducing the many-particle density matrix formalism to
single-particle one, we will also
construct an efficient approach for large-scale (e.g. molecular)
system applications.


The typical transport setup, see Fig.\ 1(A), can be described by
 \bea\label{H-ms} H &=&
 H_S(a_{\mu}^{\dg},a_{\mu})+\sum_{\alpha=L,R}\sum_{\mu k}
       \epsilon_{\alpha\mu k}d^{\dg}_{\alpha\mu k}d_{\alpha\mu k}  \nl
  & &  + \sum_{\alpha=L,R}\sum_{\mu k}(t_{\alpha\mu k}a^{\dg}_{\mu}
       d_{\alpha\mu k}+\rm{H.c.}) .
 \eea
$H_S$ is  the system Hamiltonian, which can be rather
general (e.g. including many-body interaction). $a^{\dg}_{\mu}$
($a_{\mu}$) is the creation (annihilation) operator of electron in
state labelled by ``$\mu$", which labels both the multi-orbital
and distinct spin states of the transport system.
The second and third terms describe, respectively, the two (left
and right) electrodes and the tunneling between the electrodes and
the system.

To contact with the quantum dissipation theory for
quantum open systems,
let us introduce the reservoir operators
 $F_{\mu} = \sum_{\alpha k} t_{\alpha\mu k}d_{\alpha\mu k}
  \equiv f_{L\mu} + f_{R\mu}$.
Accordingly, the tunneling Hamiltonian $H'$ reads
$ H' = \sum_{\mu} \left( a^{\dg}_{\mu} F_{\mu}
       + \rm{H.c.}\right) $.
Treating $H'$ perturbatively up to the cumulant second-order expansion,
it yields \cite{Yan98}
 \be\label{cumm-expan}
 \dot\rho(t)=-i{\cal L} \rho(t)-\int_{0}^{t} d\tau
             \la {\cal L}'(t) {\cal G}(t,\tau){\cal L}'(\tau)
             {\cal G}^{\dg}(t,\tau)\ra \rho(t).
 \ee
The reduced density matrix is defined
as $\rho(t)=\rm{Tr}_B[\rho_T(t)]$,
by tracing out
the reservoir degrees of freedom from the total
system-plus-reservoirs density matrix.
The Liouvillian superoperators are defined as
 ${\cal L}(\cdots)\equiv [H_S,(\cdots)]$,
 ${\cal L'}(\cdots)\equiv [H',(\cdots)]$,
and ${\cal G}(t,\tau)(\cdots)\equiv G(t,\tau)(\cdots)G^{\dg}(t,\tau)$
with $G(t,\tau)$ the usual propagator (Green's function)
associated with the system Hamiltonian $H_S$.

The integral kernel in \Eq{cumm-expan}, which is in the so-called
partial ordering prescription (POP) (or time-local) form \cite{Yan98},
describes the second-order tunneling self-energy.
At the second-order level,
one may replace $\rho(t)$ in the last term
of \Eq{cumm-expan} with ${\cal G}(t,\tau)\rho(\tau)$,
leading the tunneling integral kernel to
$\la {\cal L}'(t) {\cal G}(t,\tau){\cal L}'(\tau) \ra\rho(\tau)$,
being in the chronological ordering prescription
(COP) (or memory) form \cite{Yan98}.
 The corresponding four terms in the conventional Hilbert space \cite{Li04b,Li04c},
 depicted on the real-time Keldysh contour in Fig.\ 2,
 provide a clear diagrammatic interpretation for the second-order
 tunneling self-energy process.

\begin{figure}
\begin{center}
\includegraphics*[width=8cm,height=5cm,keepaspectratio]{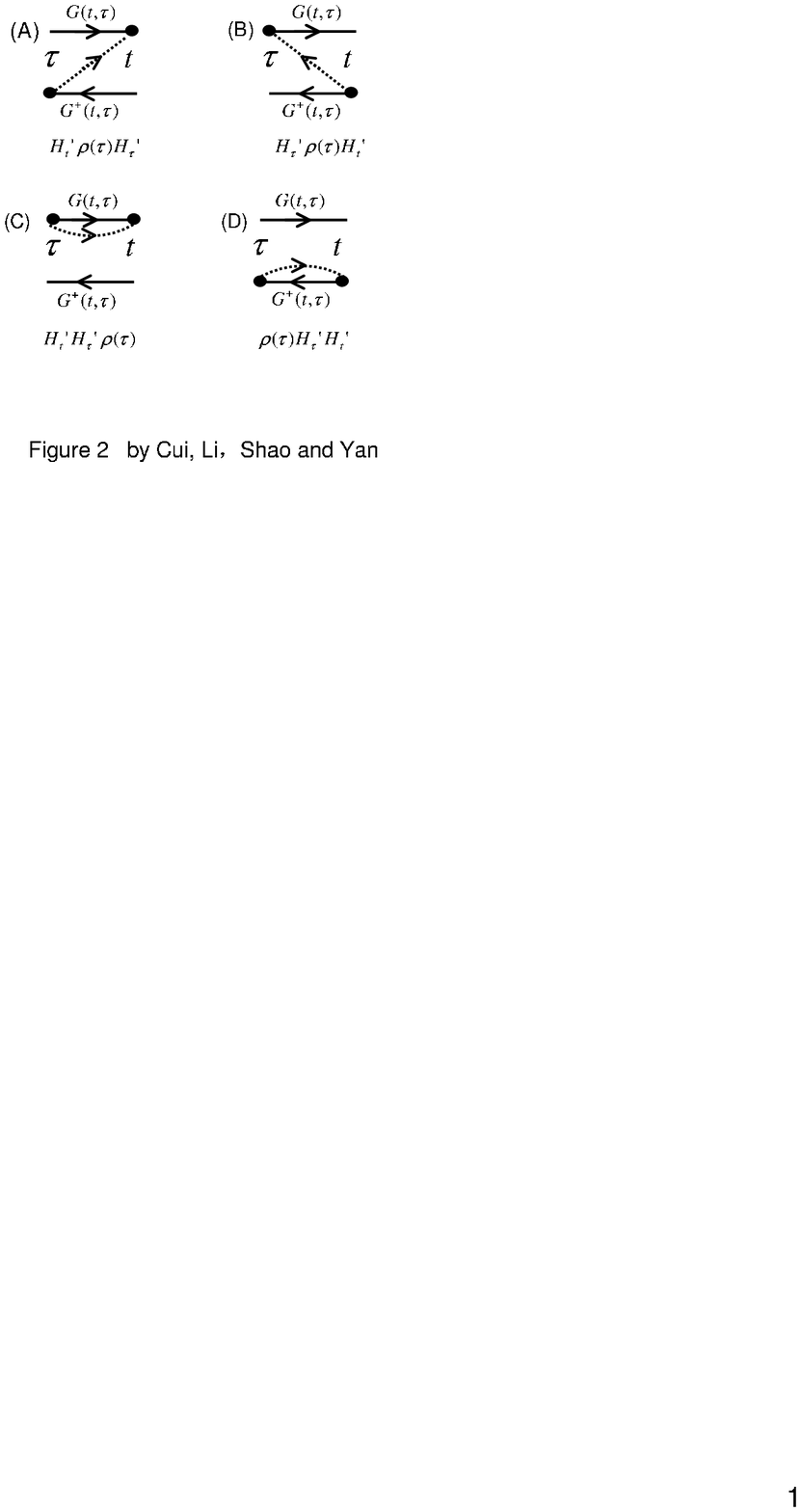}
\caption{Diagrammatic illustration for the second-order tunneling
self-energy processes, on the Keldysh contour. The upper and lower
horizontal lines describe the forward and backward propagation of
the transport system, which is treated exactly in terms of the
system Green's function $G(t,\tau)$. The dashed line stands for
the tunneling between the system and electrodes. }
\end{center}
\end{figure}

Explicitly tracing out the states of electrodes,
\Eq{cumm-expan} gives rise to
\bea\label{rho-t1}
 \dot{\rho} &=& -i {\cal L}\rho - \sum_{\mu}
        \left\{ [a_{\mu}^{\dg},A_{\mu}^{(-)}\rho
        -\rho A_{\mu}^{(+)}]  
        + \rm{H.c.} \right\}.
 \eea
For {\it time-independent} system Hamiltonian,
$ A^{(\pm)}_{\mu}=\sum_{\alpha=L,R}A^{(\pm)}_{\alpha\mu}
    = \sum_{\alpha\nu} \int_{-\infty}^{\infty}\!\!dt\,
     C^{(\pm)}_{\alpha\mu\nu}(\mp t)[i{\ti a}_{\nu}(t)] $,
with
 $C^{(+)}_{\alpha\mu\nu}(t) \equiv \la f^{\dg}_{\alpha\mu}(t)
 f_{\alpha\nu}(0)\ra$,
 $C^{(-)}_{\alpha\mu\nu}(t) \equiv \la f_{\alpha\mu}(t)
 f_{\alpha\nu}^{\dg}(0)\ra$,
 and
 ${\ti a}_{\nu}(t)=-i\Theta(t){\cal G}(t,0)a_{\nu}
\equiv \Pi^{(0)}(t,0)a_{\nu}$. Note that the step function $\Theta(t)$
extends the lower bound of the time integral from $0$ to $-\infty$,
whereas the extension of the upper bound from $t$ to $\infty$
results from the consideration of Markovian approximation.

For {\it time-dependent} system Hamiltonian,
the time-translational invariance breaks down, we thus
define ${\ti a}_{\nu}(t,t')=-i\Theta(t-t'){\cal G}(t,t')a_{\nu}$.
The backward evolution of $\ti{a}_{\nu}(t,t')$ with respect to $t'$,
starting from $t'=t$,
can be carried out via
$\partial_{t'} a_{\nu}(t,t')=i\delta(t-t')a_{\nu}
+i [H_S(t'),a_{\nu}(t,t')]$.
Thus, the time integral in $A_{\mu}^{(\pm)}$, which becomes now the type of
$\sim \int^{t}_{0}dt'C^{(\pm)}_{\alpha\mu\nu}(t-t'){\ti a}_{\nu}(t,t')$,
can be calculated accordingly.
Inserting the obtained $A_{\mu}^{(\pm)}$ into \Eq{rho-t1},
the time-dependent phenomena
associated with either quantum dissipative dynamics
or transport current can be easily treated.
For clarity, we hereafter assume the system Hamiltonian to be
time-independent, unless further specification.

Now we consider the possibility to go beyond the second-order self-energy
process diagrammatically shown in Fig.\ 1.
An efficient scheme follows the idea of the well-known
self-consistent Born approximation (SCBA), i.e.,
the free propagator defined above,
$\Pi^{(0)}(t)\equiv -i\Theta(t){\cal G}(t,0) $, is replaced
by an effective propagator $\Pi(t)$.
The latter is obtained formally via the Dyson equation
\cite{note-2}
\bea\label{pi-2}
\dot{\Pi}(t)=-i\delta(t)-i{\cal L}\Pi(t)
 -i\int^{t}_{-\infty}dt'\Sigma(t-t')\Pi(t') ,
\eea
or $\Pi(\omega) = [\omega-{\cal L}-\Sigma(\omega)]^{-1}$ in
frequency domain,
with $\Sigma$ being the irreducible self-energy defined again by
Fig.\ 2.
Accordingly, $\ti{a}_{\nu}(t)=\Pi(t)a_{\nu}$, and
\bea\label{Apm-2}
A^{(\pm)}_{\mu} = \sum_{\alpha\nu}\int\frac{d\omega}{2\pi}
           C^{(\pm)}_{\alpha\mu\nu}(\pm\omega) [i\Pi(\omega)a_{\nu}] .
\eea
Equations (\ref{rho-t1})--(\ref{Apm-2}) constitute the basic
ingredients of the proposed SCBA scheme.
This type of self-consistently partial summation correction
has included an infinite number of higher order tunneling processes
into the reduced system dynamics.
The resulting non-trivial effect on
quantum transport will be demonstrated soon.

So far, the trace is performed over all the electrode states,
and the resulting \Eq{rho-t1} is
a {\it unconditional} master equation for the system reduced dynamics.
To characterize the transport problem, we should keep track of the record
of electron numbers entering the right reservoir (electrode).
Following Refs.\ \onlinecite{Li04b} and \onlinecite{Li04c},
one can obtain
a {\it conditional} master equation for
the reduced system density matrix,
$\rho^{(n)}(t)$, under the condition that
$n$ electrons have
arrived at the right electrode until time $t$.
On the basis of $\rho^{(n)}(t)$, one is readily able to compute various
transport properties, such as the transport current, the
probability distribution function $P(n,t)\equiv\rm{Tr}[\rho^{(n)}(t)]$,
and the noise spectrum \cite{Li04b}.
For transport current, it can be carried out via
$I(t)=e \sum_n n \rm{Tr}[\dot{\rho}^{(n)}(t)$, giving rise to
\bea\label{I-t}
 I(t)   = e \sum_{\mu} \rm{Tr} \left[ \left(
         a^{\dg}_{\mu}A^{(-)}_{R\mu}-A^{(+)}_{R\mu}a^{\dg}_{\mu}
         \right)\rho(t)+\rm{H.c.} \right] .
 \eea
Compared to other transport formalisms, \Eqs{rho-t1}-(\ref{I-t})
provide a convenient framework for quantum transport.

As an illustrative application, we consider the non-trivial problem
of quantum transport through strongly interacting quantum dot,
under the well-known Anderson impurity model Hamiltonian:
$ H_S = \sum_{\mu}(\ep_0a_{\mu}^{\dg}a_{\mu}
       +\frac{U}{2}n_{\mu}n_{\bar{\mu}}) $.
Here the index $\mu$ labels the spin up (``$\uparrow$") and spin
down (``$\downarrow$") states, and $\bar{\mu}$ stands for the
opposite spin orientation. The electron number operator
$n_{\mu}=a^{\dg}_{\mu}a_{\mu}$, and the Hubbard term
$Un_{\uparrow}n_{\downarrow}$ describe the charging effect.
Apparently, the reservoir correlation function is diagonal with
respect to the spin indices, i.e.,
$C^{(\pm)}_{\alpha\mu\nu}(t)=\delta_{\mu\nu}C^{(\pm)}_{\alpha\mu\mu}(t)$,
and
 $C_{\alpha\mu\mu}^{(\pm)}(t)=\sum_k |t_{\alpha\mu k}|^2
  e^{\pm i\ep_k(t)}n^{(\pm)}_{\alpha}(\ep_k) $.
Here $n^{(+)}_{\alpha}(\ep_k)= n_{\alpha}(\ep_k)$ is the Fermi
distribution function, and $n^{(-)}_{\alpha}(\ep_k)=
1-n_{\alpha}(\ep_k)$. Accordingly, we have
$  A^{(\pm)}_{\alpha\mu}
 = \Gamma_{\alpha\mu} \int\frac{d\epsilon}{2\pi}
   n_{\alpha}^{(\pm)}(\epsilon)[i\Pi(\epsilon)]a_{\mu} $,
where, under the wide-band approximation, we have introduced
$\Gamma_{\alpha\mu}= 2\pi g_{\alpha} |t_{\alpha\mu k}|^{2}$,
and assumed it energy independent.
From \Eqs{I-t} and (\ref{rho-t1}), the stationary current is obtained as
\bea\label{IV-2}
I &=& \frac{e\Gamma_{L}\Gamma_{R}}{\Gamma_{L}+\Gamma_{R}}
    \int\frac{d\ep}{2\pi}{\rm Im}[\Pi(\ep)][n_L(\ep)-n_R(\ep)]  .
\eea
For the single level system under study, the propagator in energy space
simply reads
$\Pi(\ep)=[\ep-\ep_0-\Sigma(\ep)]^{-1}$.

Within the SCBA scheme, the self-energy $\Sigma$
can be explicitly carried out via Fig.\ 2.
However, in the case of strong Coulomb
repulsion, the dot can be occupied at most by one electron. As a
result, it can be easily proven that only Fig.\ 2(C) and (D)
contribute to the self-energy.
Physically, 
replacing the bare system propagator with the effective propagator
corresponds to including the infinitely
multiple forward and backward tunnelings between the system and the
{\it same} electrode. This is in fact a tunneling-induced quantum
fluctuation, which would lead to the level broadening
and the non-trivial interference between tunneling
and system internal interaction.
Explicitly, in large-$U$ limit, the real and imaginary parts
of the self-energy read
 $ {\rm Re} \Sigma(\ep) = (m-1)
   \sum_{\alpha=L,R}\frac{\Gamma_{\alpha\mu}}{2\pi}\Big[
   \ln\Big(\frac{\beta U}{2\pi}\Big)
   -{\rm Re}\psi\Big(\frac{1}{2}
   +i\frac{\beta}{2\pi}(\ep-\mu_{\alpha})\Big )\Big ] $
and
$ {\rm Im} \Sigma (\ep) =
   -\sum_{\alpha=L,R}\frac{\Gamma_{\alpha\mu}}{2}\Big [
   1 + (m-1)n_{\alpha}(\ep)\Big ]$,
respectively \cite{Sch94,Kon96}.
Here $\beta \equiv 1/(k_BT)$ is the inverse temperature,
$\mu_{\alpha}$ the chemical potential of the electrode,
$\psi$ the digamma function,
and $m$ denotes the spin degeneracy.
(i) For $m=1$, i.e.\ neglecting the spin degree of freedom,
${\rm Im}\Pi(\ep)$ gives the well-known Breit-Wigner formula,
which appropriately includes the level broadening effect.
(ii) For $m\ge 2$ (e.g., $m=2$ for spin $1/2$),
the above self-energy correction would result in rich behaviors,
depending on the relative values of the parameters such as
the temperature and the position of $\ep_0$ with respect to the Fermi levels.
Detailed discussions, in particular the non-equilibrium Kondo effect,
are referred to literature,
e.g.\ Refs.\ \onlinecite{Kon96}-\onlinecite{Mei93}.

\vspace{3ex} {\it Application to Large Scale Systems}.---
By far, the transport-related density matrix formalism has been
constructed in many-particle Hilbert space, which may restrict its
direct application only in small systems.
For large-scale systems in the absence of many-electron
interaction, we first recast the formalism to a very simple
version in terms of the reduced {\it single-particle} density
matrix (RSPDM), $\sigma_{\mu\nu}(t)\equiv
\rm{Tr}[a^{\dg}_{\nu}a_{\mu}\rho(t)]$, which greatly reduces the
dimension of Hilbert space, thus saves computing expense.
To account for the electron-electron interaction, we then propose
an efficient time-dependent density functional theory (TDDFT)
scheme. Note that it is quite natural to combine the TDDFT technique with the
present RSPDM formalism, since the former self-consistently
amounts to the many-body interaction but still keeps the single-particle
picture \cite{note-5}.

(i) {\it Time Independent System Hamiltonian}: For simplicity, we
first proceed our derivation in the single-particle eigenstate
basis, which is denoted as $\{|\mu\ra,|\nu\ra,\cdots\}$. In this
representation,
$A_{\alpha\mu}^{(\pm)}=
 \sum_{\nu}C^{(\pm)}_{\alpha\mu\nu}(\mp\ep_{\nu})a_{\nu}$,
and the equation of motion for the RSPDM can be readily obtained
by applying \Eq{rho-t1} directly for
$\sigma_{\mu\nu}(t)=\rm{Tr}[a^{\dg}_{\nu}a_{\mu}\rho(t)]$.
We have \cite{note-4,Yok04}
\bea\label{S-ME-2} \dot{\sigma} = -i
[h,\sigma]-\frac{1}{2} \left\{\left[C^{(-)}\sigma
  -C^{(+)}\bar{\sigma}\right]+\rm{H.c.} \right\} .
\eea
Here, $h$ is the single-particle Hamiltonian or the Fock
matrix within the TDDFT framework which will be identified soon.
$\bar{\sigma}\equiv 1 - \sigma $ denotes the ``hole'' density matrix.
The involving matrix products are defined as usual; e.g.,
$[C^{(-)}\sigma]_{\mu\nu} \equiv \sum_{\alpha=L,R}\sum_{\nu'}
C^{(-)}_{\alpha\mu\nu'}(\ep_{\nu'})\sigma_{\nu'\nu}$.
Straightforwardly, the current can be expressed in terms of the
RSPDM as
\bea\label{S-It} I(t) = e \rm{Re}\left\{\rm{Tr}\left[
C^{(-)}_{R}\sigma(t)
         -C^{(+)}_{R}\bar{\sigma}(t)\right] \right\} .
\eea
In arbitrary state basis, derivation is the same as above. The
difference lies only at the expression of
$A_{\alpha\mu}^{(\pm)}$,   
which in a non-eigenstate representation is given formally as
$A_{\alpha\mu}^{(\pm)}\equiv\sum_{\nu}\ti{C}^{(\pm)}_{\alpha\mu\nu}a_{\nu}
=\sum_{\nu\nu',m}
  C^{(\pm)}_{\alpha\mu\nu'}(\mp \ep_m) 
  D^{-1}_{\nu' m}D_{m\nu}a_{\nu} $.
Here $\ep_m$ is the eigen-energy of eigenstate $|m\ra$, and $D$ is
the transformation matrix from the non-eigestate representation to
the eigenstate one.
Obviously, with this identification, the resultant master equation
and current formula are the same as \Eqs{S-ME-2} and (\ref{S-It}),
only replacing the matrices $C^{(\pm)}$ by $\ti{C}^{(\pm)}$.

As an illustrative application of \Eqs{S-ME-2} and (\ref{S-It}),
we consider the simple non-interacting multi-level model studied
in Ref.\ \onlinecite{Li04c}. In the non-equilibrium stationary
state, $\sigma(t\rightarrow \infty)$ is diagonal in the eigenstate
basis, thus $[h,\sigma]=0$. As a consequence, the stationary state
solution is determined by
$C^{(-)}_{\mu\mu}(\ep_{\mu})\sigma_{\mu\mu}
=C^{(+)}_{\mu\mu}(-\ep_{\mu})(1-\sigma_{\mu\mu})$, leading to the
well-known result \cite{Li04c},
$\sigma_{\mu\mu}=[\Gamma_L(\ep_{\mu})n_L(\ep_{\mu})+n_R(\ep_{\mu})
\Gamma_R(\ep_{\mu})]/[\Gamma_L(\ep_{\mu})+\Gamma_R(\ep_{\mu})]$.
In particular, in the special case of equilibrium,
$\sigma_{\mu\mu}$ reduces to the Fermi-Dirac function.
Substituting $\sigma_{\mu\mu}$ into \Eq{S-It}, the well-known
resonant tunnel current is obtained.

(ii) {\it Time Dependent System Hamiltonian}:
In this case, the RSPDM can be introduced in a similar manner.
Consider, for example, 
${\rm Tr}[a^{\dg}_{\mu}A^{(-)}_{\nu}\rho(t)] = \sum_{\alpha\nu'}
\int^{t}_{0}dt' C^{(-)}_{\alpha\nu\nu'}(t,t')\sigma_{\nu'\mu}(t',t)
\equiv [C^{(-)}\sigma]_{\nu\mu}$.
Here, $\sigma_{\nu'\mu}(t',t)\equiv {\rm Tr}
\{a^{\dg}_{\mu} [{\cal G}(t,t')a_{\nu'}]\rho(t)\}$,
which can be solved via
$\partial_{t'}\sigma_{\nu'\mu}(t',t)=-i[h(t')\sigma(t',t)]_{\nu'\mu}$,
with the initial condition
$\sigma_{\nu'\mu}(t,t)={\rm Tr}[a^{\dg}_{\mu}a_{\nu'}\rho(t)]$.
Similarly, we have ${\rm Tr}[A^{(+)}_{\nu}a^{\dg}_{\mu}\rho(t)] =
\sum_{\alpha\nu'} \int^{t}_{0}dt'
C^{(+)}_{\alpha\nu\nu'}(t,t')\bar{\sigma}_{\nu'\mu}(t',t)
\equiv [C^{(+)}\bar{\sigma}]_{\nu\mu}$.
Here, $\bar{\sigma}_{\nu'\mu}(t',t)\equiv {\rm
Tr} \{[{\cal G}(t,t')a_{\nu'}]a^{\dg}_{\mu}\rho(t)\}$, satisfying
an equation of the same form as $\sigma_{\nu'\mu}(t',t)$, but with
initial condition
$\bar{\sigma}_{\nu'\mu}(t,t)=\delta_{\nu'\mu}-\sigma_{\nu'\mu}(t)$.
As a result, in the time-dependent case, the resultant master
equation and transport current can also be expressed as
\Eqs{S-ME-2} and (\ref{S-It}), only keeping in mind that the
matrices product needs not only the inner-state summation, but
also the ``inner-time'' integration.

Now we extend the above RSPDM formalism, i.e., \Eqs{S-ME-2} and
(\ref{S-It}), to interacting systems. Within the TDDFT framework
\cite{RG84}, this can be straightforwardly done by replacing the
single particle Hamiltonian by the Fock matrix \bea\label{Fock}
h_{mn}(t) = h^{0}_{mn}(t)+v^{\rm xc}_{mn}(t)
      +\sum_{ij}\sigma_{ij}(t)V_{mnij}.
\eea In first-principles calculation the state basis is usually
chosen as the local atomic orbitals, $\{ \phi_m({\bf r}), m=1,2,
\cdots \}$.
Here $h^0(t)$ is the non-interacting Hamiltonian which can be in
general time-dependent; $V_{mnij}$ is the two-electron Coulomb
integral, $V_{mnij}=\int d{\bf r}\int d{\bf r'}\phi^*_m({\bf
r})\phi_n({\bf r}) \frac{1}{|{\bf r}-{\bf r'}|}\phi^*_i({\bf
r'})\phi_j({\bf r'})$; and $v^{\rm xc}_{mn}(t)=\int d{\bf r}
\phi^*_m({\bf r})v^{\rm xc}[n]({\bf r},t)\phi_n({\bf r})$, with
$v^{\rm xc}[n]({\bf r},t)$ the exchange-correlation potential,
which is defined by the functional derivative of the the
exchange-correlation functional $A^{\rm xc}$. In practice,
especially in the time-dependent case, the unknown functional
$A^{\rm xc}$ can be approximated by the energy functional $E^{\rm
xc}$, obtained in the Kohn-Sham theory and further with the local
density approximation (LDA).
Notice that the density function $n({\bf r},t)$ appeared in the
Fock operator is related to the RSPDM via $n({\bf
r},t)=\sum_{mn}\phi_m({\bf r})\sigma_{mn}(t)\phi^*_n({\bf r})$.
Thus, \Eqs{S-ME-2}-(\ref{Fock}) constitute a closed form of TDDFT
approach for the first-principles study of quantum transport,
which is currently an intensive research subject \cite{Bur05}.

To summarize, we have proposed a compact transport formalism from
the perspective of quantum open systems. The new formulation is
constructed in terms of an improved reduced density matrix
approach at the SCBA level, which is shown to be accurate enough
in practice. Based on the established density matrix formalism, we
also developed a new TDDFT scheme for first-principles study of
transport through complex large-scale systems.
Systematic applications and numerical implementations are in
progress and will be published elsewhere.

\vspace{2ex} {\it Acknowledgments.} Support from the National
Natural Science Foundation of China and the Research Grants
Council of the Hong Kong Government is gratefully acknowledged.


\begin{references}
\bibitem{Dat95}
S. Datta, {\it Electronic Transport in Mesoscopic Systems}
(Cambridge University Press, New York, 1995).
\bibitem{Hau96}
H. Haug and A.-P. Jauho, {\it Quantum Kinetics in Transport and
Optics of Semiconductors} (Springer-Verlag Berlin, 1996).
\bibitem{Gar00}
C.W. Gardiner and P. Zoller, {\it Quantum Noise} (Springer-Verlag
Berlin, 2000).
\bibitem{Gla88}
L.I. Glazman and K.A. Matveev, JETP Lett. {\bf 48}, 445 (1988);
D.V. Averin and A.N. Korotkov, Sov. Phys. JETP {\bf 70}, 937
(1990); C.W.J. Beenakker, Phys. Rev. B {\bf 44}, 1646 (1991);
Yu.V. Nazarov, Physica B {\bf 189}, 57 (1993); S.A. Gurvitz, H.J.
Lipkin, and Ya.S. Prager, Phys. Lett. A {\bf 212}, 91 (1996).
\bibitem{Gur96}
S.A. Gurvitz and Ya.S. Prager, Phys. Rev. B {\bf 53}, 15932
(1996).
\bibitem{Li04a}
X.Q. Li, W.K. Zhang, P. Cui, J.S. Shao, Z.S. Ma, and Y.J. Yan,
Phys. Rev. B {\bf 69}, 085315 (2004) (e-print cond-mat/0309574).
\bibitem{Li04b}
X.Q. Li, P. Cui, and Y.J. Yan, Phys. Rev. Lett. {\bf 94}, 066803
(2005) (e-print quant-ph/0408073).
\bibitem{Li04c}
X.Q. Li, J.Y. Luo, Y.G. Yang, P. Cui, and Y.J. Yan, Phys. Rev. B
{\bf 71}, 205304 (2005).
\bibitem{Yan98}
Y.J. Yan, Phys. Rev. A {\bf 58}, 2721 (1998).
\bibitem{note-2}
Here, the usual Dyson equation has been converted into its
equivalent equation of motion in ``COP" form.
\bibitem{Kon96}
J. K\"{o}nig, J. Schmid, H. Scheoller, and G. Sch\"{o}n, Phys.
Rev. B {\bf 54}, 16820, (1996).
\bibitem{Sch94}
H. Schoeller and G. Sch\"{o}n, Phys. Rev. B {\bf 50}, 18436,
(1994).
\bibitem{Mei93}
Y. Meir, N.S. Wingreen, and P.A. Lee, Phys. Rev. Lett. {\bf 70},
2601 (1993).
\bibitem{note-5}
To make the scheme at the level of first-principles, the transport system
(i.e. the device) can be extended to include part of the electrodes.
This {\it extended device} will be treated by using the quantum chemical approach
such as the DFT, meanwhile the effect of the other part of the electrodes
is described by a self-energy matrix ``$\Sigma$" that can be computed
from the surface Green's
function technique. The broadening matrix ``$\Gamma$", which appears in the
correlation function $C^{(\pm)}$, is then obtained by $\Gamma=i(\Sigma-\Sigma^{\dg})$.
\bibitem{note-4}
For simplicity, in what follows our description is based on the
Born approximation. Generalization to the SCBA is straightforward.
As demonstrated in Ref.\ \onlinecite{Li04c},
in most applications the Born approximation will be already good enough.
\bibitem{Yok04}
S. Yokojima, G.H. Chen, R.X. Xu, and Y. J. Yan, Chem. Phys. Lett.
{\bf 369}, 495 (2004).
\bibitem{RG84}
E. Runge and E.K.U. Gross, Phys. Rev. Lett. {\bf 52},997 (1984);
C.Y. Yam, S.Yokojima, and G.H. Chen, J. Chem. Phys. {\bf 119},
8794 (2003); Phys. Rev. B {\bf 68}, 153105 (2003).
\bibitem{Bur05}
M.A.L. Marques and E.K.U. Gross, Annu. Rev. Phys. Chem. {\bf 55},
427 (2004); K. Burke, R. Car, and R. Gebauer, Phys. Rev. Lett.
{\bf 94}, 146803 (2005);
X. Zheng and G.H. Chen (unpublished).
\end{references}
\end{document}